\documentclass[12pt,a4paper]{article} 

\pdfoutput=1
\usepackage[numbers]{natbib}
\usepackage[affil-it]{authblk}

\usepackage{amsmath}
\usepackage{amsthm}
\usepackage[english]{babel}
\usepackage{graphicx}
\usepackage{amsfonts} 
\usepackage{amssymb} 
\usepackage{hyperref} 
\usepackage{float}
\usepackage{times}
\usepackage{tensor}
\usepackage{mathtools}
\usepackage{color}
\usepackage{bbold}

\usepackage{subdepth}
\usepackage{mathrsfs}

\newcommand{\pd}[1]{\frac{\partial}{\partial #1}}
\newcommand{\PD}[2]{\frac{\partial #1}{\partial #2}}
\newcommand{\fd}[1]{\frac{\delta}{\delta #1}}

\newcommand{\de}{\partial}
\newcommand{\lqu}{\textrm`}
\newcommand{\rqu}{\textrm'}
\newcommand{\QU}[1]{[\lqu #1 \rqu]}

\newcommand{\oppr}[3]{\ensuremath{\left \langle \left. #1\right.  \right| #2 \left| \left. #3 \right. \right \rangle}}

\newcommand{\gtilde}	{\tilde{g}}
\newcommand{\pitilde}	{\tilde{\pi}}
\newcommand{\gdot}	{\dot{g}}
\newcommand{\gbar}	{\bar{g}}

\newcommand{\phibar}	{\bar{\phi}}
\newcommand{\phidot}	{\dot{\phi}}

\newcommand{\pphi}[1]	{{p_{\phi_{#1}}}}
\newcommand{\pphibar}[1]{{\bar{p}}_{\phi_{#1}}}
\newcommand{\pphisq}	{p_\phi^2}

\newcommand{\rootg}	{\sqrt{g}}

\newcommand{\tk}	{2\kappa}
\newcommand{\curlyH}	{\mathcal{H}}
\newcommand{\zero}	{{(0)}}
\newcommand{\one}	{{(1)}}
\newcommand{\two}	{{(2)}}

\newcommand{\Rbar}	{\bar{R}}

\title{Cosmological perturbation theory with York time}
\author{Philipp Roser\thanks{proser@clemson.edu}}
\affil{Department of Physics and Astronomy,\\ Clemson University, Kinard Laboratory,\\ Clemson, SC 29631-0978, USA}
\date{\vspace{-0.5cm}}		%ELIMINATES DATE

\begin{document}
 \maketitle
 %\tableofcontents
%------------------------------------
\begin{abstract}
 One method to overcome the notorious problem of time in the quantisation of gravity is the identification of a physically preferred time parameter, a promising candidate being so-called `York time'. The dynamical equations for matter and spatial geometry in York time may be obtained via Hamiltonian reduction, that is, by solving the Hamiltonian constraint for the physical, non-vanishing Hamiltonian density identified as the variable conjugate to the chosen time parameter. Yet in general this equation cannot be solved algebraically. Here we show how in a cosmological scenario, where one may treat geometric and matter inhomogeneities as small perturbations, one is able to obtain the physical Hamiltonian density by solving the constraint equation perturbatively. By construction the Hamiltonian density is quadratic in the perturbation variables, which makes it easily quantisable, although subtleties arise due to the non-canonical form of the Poisson brackets and the time-dependent coefficients. The latter are determined by the evolution of the background variables. %Since in quantisation `time' is treated as a parameter rather than a physical variable, different choices of physical time may lead to phenomenologically distinct quantum theories, allowing in principle the empirical identification of the correct choice of physical time. 
\end{abstract}
\newpage
%_________________________________________________________________________________

\section{Introduction} \label{Introduction}

The elusiveness of a complete theory of quantum gravity is at least in part due to the difficulty of performing a na\"ive quantisation of general relativity in the manner analogous to other gauge theories. This difficulty is referred to as `the problem of time' and has multiple facets \citep{Kuchar2011,Anderson2012}, perhaps the most prominent of which is the `frozen' dynamics one obtains when imposing the ADM Hamiltonian constraint \citep{ADM1962} on the quantum level \citep{DeWitt1967,UnruhWald1989, Rovelli2007}. The problems arise, roughly speaking, due to the arbitrariness of the space-time foliation chosen for quantisation and the frozen dynamics in particular follows from the reparameterisation invariance of the classical theory even once the foliation is fixed. 

A substantial literature has been dedicated to a large variety of ideas aimed at overcoming this problem. One proposal is to break the refoliation and reparameterisation invariance of the theory at the classical level by selecting a physically preferred time parameter. Associated with this time parameter one would have a Hamiltonian, which unlike the Hamiltonian in the reparameterisation-invariant theory, is non-vanishing and therefore does not lead to a frozen dynamics. However, while this idea may in principle be straightforward, it requires one to first identify this physically preferred time parameter in the classical theory. The parameter is presumably to be chosen from among the physical variables in the classical theory and as such may be extrinsic, that is derived from the gravitational/geometric degrees of freedom including the extrinsic curvature, intrinsic, that is chosen from among the gravitational/geometric 3-quantities alone, or extracted from the matter content of one's theory (see e.g.\ \citep{BrownKuchar1995,HusainPawlowski2012} for an example of the latter).

One proposal of interest is the identification of the scalar extrinsic curvature of a spatial slice as the time parameter. The primary reason for this choice is the role of the extrinsic curvature in York's solution to the initial-value problem of general relativity \citep{York1972,ChoquetBruhatYork1980} and the parameter (after a convenient rescaling by a constant) is therefore called `York time', defined as $T=\frac{2\pi}{3\sqrt{g}}$, where $\pi=g_{ab}\pi^{ab}$ is the trace of the gravitational momentum $\pi^{ab}$ and $g=\det(g_{ab})$ is the spatial 3-metric determinant. We use units such that $c=1$. More intuitively, York time is given by the local rate of fractional spatial constraction. York showed \citep{York1972, York1971, York1973} that a complete set of Cauchy data may be given on a slice of constant extrinsic curvature by first specifying $g_{ab}$ and $\pi^{ab}$ such that they satisfy the momentum constraints, and then rescaling them by a local conformal factor such that the Hamiltonian constraint is satisfied. The scale chosen initially is thereby irrelevant for the final result, strongly suggesting that the physical degrees of freedom are the scale-free parts of the metric and its associated momentum. In other words, the initial-value problem is solved on slices of constant York time $T$, and $T$ and its conjugate momentum $P_T$ are thereby shown to be not among the natural physical variables. 

In recent years the the initial-value problem together with advances in relational theories \citep{BarbourFosterOMurchadha2002, AndersonBarbourFosterOMurchadha2003, AndersonBarbourFosterKelleherOMurchadha2005} have led to the theory of \emph{shape dynamics} (\citep{GomesGrybKoslowski2011,BarbourKoslowskiMercati2013,Mercati2014} and references in \citep{Mercati2014}), an alternative theory of gravity where the refoliation symmetry of general relativity is traded for a local conformal symmetry. The emerging time parameter in this theory is York time rescaled to be dimensionless by taking its ratio with an arbitrary reference time. Shape dynamics and general relativity are equivalent for space-times that allow a globally hyperbolic foliation but may differ in their predictions if no such foliation exists, for example, in the presence of black hole singularities \citep{Gomes2014,GomesHerczeg2014}. The existence of shape dynamics and its relationship with York time is another reason to explore the latter, although in this paper we will be concerned with a reduced-Hamiltonian theory derived directly from general relativity and not shape dynamics.

A reason from quantum theory, in particular hidden-variable approaches, to consider York time as physically preferred is discussed in \citep{Valentini1996}. Some properties of constant-mean-curvature foliations which are advantageous to ascribing it physical significance are found in \citep{QadirWheeler1985}.

Having chosen a physical time parameter $T$ one can obtain an associated Hamiltonian density $\curlyH$ by solving the Hamiltonian constraint for the momentum $P_T$ conjugate to $T$ in terms of the other dynamical variables and then defining $\curlyH\equiv-P_T$. This process is referred to as Hamiltonian reduction and the dynamical variables `left over' after a time parameter and associated momentum have been extracted are called the reduced variables. We performed a Hamiltonian reduction for the Friedmann universe together with a scalar field and discussed properties of the resulting quantum theory in \citep{RoserValentini2014a}. Furthermore, depending on the choice of $T$ the reduced variables may be non-canonical as evident from their Poisson structure. This is the case for York time as we will discuss in more detail in section \ref{PertPrinciples}. We explored some of the implication of this Poisson structure, in particular for quantisation, via an anisotropic minisuperspace model in \citep{Roser2015a}.

However, solving the Hamiltonian constraint analytically is not possible in general \citep{ChoquetBruhatYork1980}. In this paper we instead aim to solve the constraint perturbatively and derive a classical and quantum cosmological perturbation theory with York time as the physical time parameter. It is noteworthy that, while on the classical level the choice of time parameter may be seen as purely aesthetic --- that is, any such reduced-Hamiltonian theory will agree with the predictions of general relativity --- this is not necessarily so on the quantum level. This is because time is not treated as a dynamical variable during quantisation. Distinct choices of time parameter imply distinct sets of reduced variables and the resulting quantisation of different sets of variables may, in general, lead to different phenomenologies. We leave the details of these phenomenological differences to future work. Here we merely wish to emphasise that the choice of physical time parameter is not merely a choice of `coordinates'.

\section{Principles of perturbation theory in York time} \label{PertPrinciples}

Solving the Hamiltonian constraint perturbatively does, of course, limit the applicability of the reduced-Hamiltonian theory to perturbative regimes. Fortunately we know that the observed universe is well described in terms of a homogeneous isotropic `background' together with perturbations described by linearised field equations. While the reduced Hamiltonian at which we arrive is therefore not fundamental and the quantised theory does not constitute a fundamental theory of quantum gravity, the theory is nonetheless applicable to the observed universe.

The method we employ here in order to derive the cosmological York-time perturbation theory is as follows. One first splits the geometric and matter variables and momenta appearing in the Hamiltonian constraint into homogeneous isotropic background variables, and perturbations. For matter content we choose a set of scalar fields $\phi_a$ for simplicity. One can then solve the Hamiltonian constraint to zeroth order and derive a Hamiltonian for the background dynamics, essentially the result found in \citep{RoserValentini2014a}. The zeroth-order equations of motion may then be solved and one substitutes their solutions as functions of York time back into the Hamiltonian constraint, which one can then solve to second order in the perturbation variables in order to arrive at a reduced Hamiltonian that is second order in the perturbation variables. %In practice the two steps are combined into one, however; one solves the Hamiltonian constraint to second order and only then extracts the zeroth-order dynamics before proceeding with the perturbative analysis. 
Note that first-order quantities in general vanish if the background equations of motion are satisfied, so that the Hamiltonian describing the perturbations is second order only and therefore gives linear equations of motion.

In general the reduced variables for York time are given by \citep{ChoquetBruhatYork1980}
\begin{equation}\label{generalreducedvariables}
 \gtilde_{ab} = g^{-\frac13}g_{ab},\qquad \pitilde^{ab}=g^\frac13\left(\pi^{ab}-\tfrac13\pi g^{ab}\right)
\end{equation}
and their Poisson brackets are
\begin{align}
 \{\gtilde_{ab}(\vec{x}),\pitilde^{cd}(\vec{y})\}  &= \left(\delta_a^{(c}\delta^{d)}_b-\tfrac13\gtilde_{ab}\gtilde^{cd}\right)\delta^3(\vec{x}-\vec{y}),  \label{generalgpiPB} \\
 \{\pitilde^{ab}(\vec{x}),\pitilde^{cd}(\vec{y})\} &= \frac13\left(\gtilde^{cd}\pitilde^{ab}-\gtilde^{ab}\pitilde^{cd}\right)\delta^3(\vec{x}-\vec{y}).  \label{generalpipiPB} 
\end{align}
Other Poisson brackets vanish. The non-canonicity is apparent from the existence of the second term in the parentheses in \ref{generalgpiPB} and the fact that the right-hand side of \ref{generalpipiPB} is non-vanishing. The fact that the variables are not canonical does not hinder quantisation, as illustrated in \citep{Roser2015a}. However, the fact that the momentum-momentum bracket does not vanish implies that the quantum theory cannot possess a momentum but only a position representation.

Before we can write down the perturbative expansion of the dynamical variables and the resulting Poisson structure explicitly, we must briefly discuss the question of gauge choice. The issue is well known from conventional cosmological perturbation theory \citep{MukhanovFeldmanBrandenberger1992,Mukhanov2005}: writing down some perturbative expansion of the 3-metric and matter fields does not by itself separate truly physical perturbations from apparent perturbations due to the choice of coordinate system. Furthermore, small adjustments in the coordinate system (the `gauge choice') allow one to move the physical content of the perturbation expansion between the different variables. Ultimately one is interested in the two physical degrees of freedom, manifest in gauge-independent quantities such as the Bardeen variables \citep{Bardeen1980}.

When performing perturbation theory with York time the situation is similar, although the gauge freedom is limited to 3-diffeomorphisms on the spatial slice since the foliation itself is fixed. To see this, one may begin with a York slicing at zeroth order only. In the case of a homogeneous isotropic backgroumd this foliation is exactly that of the standard Friedmann description, although the slices are differently parameterised, and one has the full gauge freedom of standard cosmological perturbation theory available. There is no bar to developing perturbation theory in this framework. However, if York time is physically preferred we ought to also impose the slicing at the level of the perturbations. This imposes a partial gauge choice, which we will refer to as the `York gauge'. From the definition of York time and its conjugate momentum one can verify that these condition effectively are that the change in $\det(g_{ab})$ and in $\pi\equiv Tr(\pi^{ab})$ are zero, or equivalently that the linear change in the metric, $\delta g_{ab}$, and in the momentum, $\delta \pi^{ab}$, are both traceless at first order. With the slicing fixed to the York gauge, one is only left with transformations of the metric and associated geometric momentum perturbation due to 3-diffeomorphisms in the surface. Conveniently the York gauge is not only the correct gauge to use from a physical standpoint, but also provides significant algebraic simplification of various terms in the perturbation expansion of the Hamiltonian constraint.

Working in the York gauge from the beginning, we perturbatively expand the reduced variables $\gtilde_{ab}$, $\pitilde^{ab}$ only rather than the original canonical variables $g_{ab}$, $\pi^{ab}$, and the variables describing the scalar field. Throughout this paper we denote zeroth-order (background) terms by an overbar. We define the perturbation variables $h_{ab}(x)$, $\nu^{ab}(x)$ in terms of the reduced variables (eq.\ \ref{generalreducedvariables}) as
\begin{align}
 \gtilde_{ab}&\equiv\gamma_{ab}+h_{ab}, \label{defofh}, \\
 \pitilde^{ab}&\equiv\tilde{\bar{\pi}}^{ab} + \nu^{ab} = \nu^{ab} \label{defofnu} ,
\end{align}
where $\gamma_{ab}$ is the scale-free metric of the background (in the case of a flat universe this is just the identity) and the last equality holds because the scale-free part of the background does not evolve, so its conjugate momentum $\tilde{\bar{\pi}}^{ab}$ vanishes. That is, the reduced momentum is perturbation only. The associated Poisson brackets to first order are
\begin{align} 
\{h_{ab}(x),\nu^{cd}(y)\} &= \Big[ \delta_a^{(c}\delta_b^{d)} - \tfrac13\gamma_{ab}\gamma^{cd} + \tfrac13\gamma^{cd}h_{ab} \notag\\
      &\hspace{0.2\linewidth}-\tfrac13\gamma_{ab}h^{cd}  +\text{h.o.}\Big]\;\delta^3(x-y), \label{hnuPB} \\
\{\nu^{ab}(x),\nu^{cd}(y)\} &= \frac13\left[\gamma^{cd}\nu^{ab}-\gamma^{ab}\nu^{cd}\right]\;\delta^3(x-y). \label{nunuPB}
\end{align}
Indices of $h_{ab}$ and $\nu^{ab}$ are raised and lowered by $\gamma^{bc}$ and $\gamma_{bc}$ respectively. These equations are derived from \ref{generalgpiPB}, \ref{generalpipiPB} and we used the fact that the expansion of the inverse metric is (to first order) $\gtilde^{ab}=\gamma^{ab}-h^{ab}$, with the scale-free inverse background metric $\gamma^{ab}$ defined via $\delta_a^c=\gamma_{ab}\gamma^{bc}$. %The expansion of this and other useful quantities to second order are given in the appendix \ref{usefulformulae}.

By construction the physical Hamiltonian obtained in the reduction is quad\-ratic because the perturbation expansion is to second order, although the terms contain time-depedent coefficients determined by the solution of the background dynamics. Note that both background and perturbations will be quantised. In order to obtain trajectories for the background dynamics, that is, functions of time which will constitute the coefficients of the perturbation terms, we employ the de~Broglie-Bohm pilot-wave formulation of quantum mechanics \citep{deBroglie1928,Bohm1952,Holland1993}, where such trajectories are part of the fundamental ontology. De~Broglie-Bohm trajectories have previously been used as a mathematical tool in cosmological perturbation theory in \citep{Pinto-NetoSantosStruyve2012,Pinto-NetoSantosStruyve2014}, although with conventional cosmological time rather than York time. The fact that the Hamiltonian is a sum of quadratic terms means that the quantisation procedure is straightforward once a representation of the commutator algebra derived from the non-canonical Poisson structure is found. The time-dependent coefficients do however make finding solutions to the quantum dynamics more difficult. 

We stated above that in general the reduced variables do not allow for a momentum representation due to the non-vanishing of the right-hand side of eq.\ \ref{generalpipiPB}, since such a representation in the quantum theory would require that for a momentum wave-functional $\tilde{\Psi}$ one has 
\begin{equation} \label{Pidontcommute}
\hat{\pi}^{ab}\hat{\pi}^{cd}\tilde{\Psi}[\pi^{ij}(x)]=\pi^{ab}\pi^{cd}\tilde{\Psi}[\pi^{ij}(x)]
	  =\pi^{cd}\pi^{ab}\tilde{\Psi}[\pi^{ij}(x)]=\hat{\pi}^{cd}\hat{\pi}^{ab}\tilde{\Psi}[\pi^{ij}(x)] ,
\end{equation}
which contradicts the commutator bracket obtained from the quantised form of eq.\ \ref{generalpipiPB}. This holds also true for the perturbative form \ref{nunuPB}, although here we note that the right-hand side vanishes at zeroth order. This implies that one can find an `approximate' momentum representation. Note on the other hand that the leading-order contribution to the non-canonical part of eq.\ \ref{hnuPB} is $-\frac13\gamma_{ab}\gamma^{cd}$, which is zeroth order.

The fact that the quantum theory has a preferred basis, the position representation, may be taken to provide hints for foundational questions in quantum theory. For example, formulations in which the position representation has a special status, such as the de~Broglie-Bohm formulation employed below, do not have to explain why the position representation should be taken as fundamental. However, these questions do not concern us for the purposes of this paper.

\section{The Hamiltonian constraint equation in York-time perturbation theory}

As outlined above, we begin with the ADM action \citep{ADM1962} for general relativity minimally coupled to a set of scalar fields $\phi_A$ with momenta $\pphi{A}$,
\begin{equation} \label{ADMaction}
 S = \int dt\,d^3x\;\left[\gdot_{ab}\pi^{ab}+\phidot_A\pphi{A} -N_i\mathscr{H}^i-N\mathscr{H}\right],
\end{equation}
where 
\begin{align}
 \mathscr{H}&= -\frac{\rootg}{\tk}R+\frac{\tk}{\rootg}\,(g_{ac}g_{bd}-\tfrac12 g_{ab}g_{cd})\pi^{ab}\pi^{cd} \notag\\
		 &\hspace{0.2\linewidth}+ \frac{1}{2\rootg}\pphisq+\frac{\rootg}{2}g^{ab}\phi_{A,a}\phi_{A,b} + \rootg V(\phi), \label{GeneralHamiltonianConstraint} \\
 \mathscr{H}^a &= -2\nabla_b\pi^{ab} + \pphi{A} g^{ab}\de_b\phi_A \label{GeneralMomentumConstraint}
\end{align}
constitute the Hamiltonian and momentum constraints respectively. Summation over repeated field indices ($A$, etc.) is assumed. Here $R$ is the scalar 3-curvature, $\tk=16\pi G$, $\pphisq=\sum_A\pphi{A}^2$ and $V(\phi)$ is a currently arbitrary potential of the scalar fields $\phi=\{\phi_1,\phi_2,\dots\}$. We suppress the spatial and temporal arguments of the field quantities where they are unambiguous in order to avoid notational clutter.

The goal is to solve the Hamiltonian constraint, $\mathscr{H}=0$, for $P_T = -\sqrt{g}$, the momentum conjugate to the York time parameter $T\equiv\frac{2\pi}{3\rootg}$. One therefore performs the change of variables $\{g_{ab},\pi^{ab}\}\rightarrow\{\gtilde_{ab},\pitilde^{ab},T,P_T\}$ \citep{ChoquetBruhatYork1980}. In the perturbative case considered here we instead make the change $\{g_{ab},\pi^{ab}\}\rightarrow\{h_{ab},\nu^{ab},T,P_T\}$ with $h_{ab}$ and $\nu^{ab}$ defined by eqs.\ \ref{defofh} and \ref{defofnu} respectively. The fields are expanded as
\begin{equation} \phi_A=\phibar_A +\delta\phi_A,\qquad\qquad\pphi{A}=\pphibar{A}+\delta\pphi{A}.\end{equation}
We keep terms up to second order in the geometric and matter perturbation variables. While not difficult, this leads to some lengthy expressions.

The scaling of the matter terms in \ref{GeneralHamiltonianConstraint} is straightforward since their only scale-dependence is in the prefactor $g^\frac12\sim P_T$ or $g^{-\frac12}\sim P_T^{-1}$. The curvature terms obtained in the expansion of $R$ are homogeneous in their scale dependence, $R\sim g^{-\frac13}\sim P_T^{-\frac23}$ and only the momentum terms are inhomogeneous in their $T,P_T$-dependence. The latter fact follows from the expression obtained when writing $\pi^{ab}$ in terms of $\tilde{\pi}^{ab}$,
\begin{align}\pi^{ab} &= g^{-\frac13}\pitilde^{ab}+\tfrac13\pi g^{-\frac13}\gtilde^{ab} \notag\\
		      &= \curlyH^{-\frac23}\nu^{ab}+\tfrac13\left(\tfrac32T\curlyH\right)\cdot\curlyH^{-\frac23}
			   \left(\gamma^{ab}-\gamma^{ac}\gamma^{bd}h_{cd}+\gamma^{ac}\gamma^{de}\gamma^{fb}h_{cd}h_{ef}\right),
\end{align} 
the last equality holding to second order. With the further identification of the physical Hamiltonian density $\curlyH\equiv-P_T$, the Hamiltonian constraint equation becomes the equation determining $\curlyH$,
\begin{align}\label{HCEq}
 0 &= \curlyH\Big(-\tfrac38\cdot\tk T^2+\QU{V}\Big) + \curlyH^\frac13\Big(\QU{R}+\QU{\nabla\phi}\Big) +\tk Th_{ab}\nu^{ab} \notag\\ 
	 &\hspace{0.4\linewidth} +\curlyH^{-1}\Big(\tk\gamma_{ac}\gamma_{bd}\nu^{ab}\nu^{cd}+\QU{p_\phi}\Big),
\end{align}
where we have introduced the shorthand $\QU{X}$ to denote the terms in the perturbative expansion after factoring out the scale dependence which are derived from the term containing $X$ in the Hamiltonian constraint \ref{GeneralHamiltonianConstraint}. With the York gauge applied these are given by
\begin{align}
 \QU{R} &= -\frac1\tk\tilde{\Rbar}-\frac1\tk\widetilde{\delta R}^\one -\frac1\tk\widetilde{\delta R}^\two  \label{Rterms}\\
 \QU{p_\phi} &= \tfrac12\bar{p}_\phi^2+\pphibar{A}\delta\pphi{A}+\delta\pphisq. \label{pphiterms}\\ 
 \QU{\nabla\phi} &= \tfrac12\gamma^{ij}\delta\phi_{A,i}\delta\phi_{A,j} \label{delphiterms}\\
 \QU{V} &= V(\phibar)+\delta\phi_A\left.\PD{V}{\phi_A}\right|_{\phibar} 
		    +\tfrac12\delta\phi_A\delta\phi_B\left.\frac{\partial^2V}{\partial\phi_A\partial\phi_B}\right|_{\phibar}.\label{Vterms}
\end{align}
The algebraic details of the expansion as well as the form of the first and second-order curvature-perturbation terms $\widetilde{\delta R}^\one$, $\widetilde{\delta R}^\two$, see appendix \ref{usefulformulae}. Throughout this paper we use superscripts `$(n)$' to denote the $n$th-order contribution to the preceeding quantity. At this point we have retained the first-order terms, although these will cancel when the background equations of motion are satisfied. This is a general result following from the fact that the equations are derivable via an extremisation principle.

%Use perturbation theory to overcome unsolvability by perturbative expansion --- but limited for perturbative regimes --- but applicable to observed universe \\
%Details of method --- role of 0th order terms as background dynamics, now functions of $T$ --- 1st order vanishes, 2nd order Hamil leads to 1st order equations \\
%PBs of perturbation variables, implications \\
%Quantisation: quadratic H --> easy --- but time dependence --- PBs imply no momentum rep, though have approximate momentum rep since $\nu$-$\nu$ bracket non-canonicity of higher order \\
%Detect physical time coordinate by finding empirically correct quantum theory? \\

%ADM action, Hamiltonian constraint $\mathfrak{H}$,$\QU{x}$ terms\\ 
%Re-introduce $h_{ij}$. $\nu^{ij}$, $\delta\phi_A$, $\delta\pphi{A}$ (used in section above)\\
%$\pi^{ij}$ in terms of $\nu^{ij}$ \\
%Vanishing of first-order terms if background on-shell (show in generality) \\

\section{Hamiltonian reduction and the equations of motion}

After multiplication by $\curlyH$ eq.\ \ref{HCEq} is a sextic equation in $\curlyH^\frac13$, which in general cannot be solved analytically. As previously anticipated, the procedure now is to first solve this equation at zeroth order and use the solution as the basis for solving it up to second order perturbatively. At each order only a new linear equation has to be solved. Since the background is homogeneous the zeroth-order terms have no spatial dependence and the equation is a simple polynomial,
\begin{equation} 0=\Big(-\tfrac38\cdot\tk T^2+V(\phibar)\Big)\curlyH^{\zero2}-\frac{1}{\tk}\tilde{\Rbar}\curlyH^{\zero\frac43}+\tfrac12p_\phi^2 .\label{BGHCEq}\end{equation}
Since the $\curlyH^0$ term in \ref{HCEq} was second order only, the zeroth-order equation does not have a corresponding term and therefore after multiplication by $\curlyH^{-2}$ takes the form of a depressed cubic in $u\equiv\curlyH^{\zero-\frac23}$, which has solutions
\begin{equation} u=(A+\sqrt{A^2-C^3})^\frac13 + (A+\sqrt{A^2-C^3})^\frac13,\end{equation}
with %DOULBE CHECK FACTORS
\begin{equation} A= \frac{1}{\pphisq}\left(\tfrac38\cdot\tk T^2-V(\phibar)\right), \qquad C= \frac{1}{6\pphisq}\frac{1}{\tk}\tilde{\Rbar} .\end{equation}
For the purposes of our model we now make the assumption that the universe is flat at zeroth order. This choice is motivated by two facts: First, the algebra in what follows is significantly less convoluted, even at the level of the background. Second, observation suggests that the universe is flat on sufficiently large scales within experimental uncertainty, giving for example (WMAP, \citep{WMAP2007}) $\Omega_k=-0.011\pm0.012$ or $-0.014\pm0.017$ depending on the exact choice of data and dark-energy model, so the choice is empirically plausible. Having chosen the global geometry of the background we can also fix the background frame of reference, making the obvious (inertial) choice $\gamma_{ab}=diag(1,1,1)$. However, for the most part we will retain reference to $\gamma_{ab}$ explicitly.

With the assumption of flatness one has $\tilde{\Rbar}=0$ and the background equation \ref{BGHCEq} is trivially solved,
\begin{equation} \curlyH^\zero = \pm\left[\frac{\frac12\bar{p}_\phi^2}{\frac38\cdot\tk T^2-V(\phibar)}\right]^\frac12. \label{flatBGHCEq}\end{equation}
The ambiguous sign has no physical effect. A change in the choice of the sign in \ref{flatBGHCEq} leads to the same set of physical trajectories with the exception that the sign of corresponding momenta are swapped. That is, the set of solutions of the dynamical equations obtained from the Hamiltonian density $\curlyH$ in \ref{flatBGHCEq} with a positive sign is related to the set of solutions of $\curlyH$ with a negative sign via a reflection in phase space. The physical interpretation of the numeric value of the Hamiltonian is that of volume (since $\curlyH=\rootg$, see \citep{RoserValentini2014a}), which suggests that $\curlyH\geq0$ is the more physically meaningful choice, and indeed we will adopt this convention here. %OTHER OPTION MATHEMATICALLY VIABLE

For a single scalar field, the dynamics of this spatially constant Hamiltonian density was discussed in \citep{RoserValentini2014a} (though note the minor correction provided in \citep{Roser2015CosmExtension}), where integration over a comoving normalisation volume was assumed. The details depend on the choice of $V(\phi)$. Note that there are no remaining geometric degrees of freedom at the background level since the scale variable $g$ and its conjugate momentum $\pi$ have been eliminated in favour of $T$ and $P_T=-\curlyH$. In the case of a set of free fields the momenta $\pphi{A}$ are constant and the fields simply evolve according to the Hamiltonian density \ref{flatBGHCEq} for $V=0$, leading to
\begin{equation} \phi_A^\prime = \frac{\pphi{A}}{\sqrt{\frac34\cdot\tk\cdot\pphisq T^2}},\end{equation}
so that, for an expanding universe ($T<0$), 
\begin{equation} \phi_A(T) = \frac{\pphi{A}}{\sqrt{\frac34\cdot\tk\cdot\pphisq}}\ln|T|.\end{equation}
The volume of the universe is the numerical value of the background Hamiltonian density obtained by substituting the solution back into expression \ref{flatBGHCEq}, up to a constant determined by the coordinate volume chosen for normalisation in the case of a flat universe as discussed above (the volume is well-defined without normalisation for a closed universe, though here the Hamiltonian takes a more complicated form).

For general $V(\phi)$ one obtains solutions ($\phi_A(T),\pphi{A}(T)$) by an appropriate method of solving Hamilton's equation for the Hamiltonian obtained after integrating $\curlyH^\zero$ over a (comoving) normalisation volume. Assuming Hamilton's equations for the background to be satisfied one can show that the first-order terms in eq.\ \ref{BGHCEq} (note that these are matter terms only) cancel out. 

With $\curlyH^\zero$ found, one can substitute the result back into eq.\ \ref{HCEq} and proceed to first order. This is trivially $[\curlyH^\frac13]^\one=0$ since there are no first-order terms. The second order equation may be solved to yield the $[\curlyH^\frac13]^\two$ from which one then obtains the second-order contribution to the Hamiltonian density, 
\begin{align} \curlyH^\two 
 &=F(T)\cdot\Bigg[ \tk\gamma_{ik}\gamma_{jl}\nu^{ij}\nu^{kl} + \curlyH^\zero\cdot\tk Th_{ij}\nu^{ij} \notag\\
	&\hspace{0.15\linewidth} -\curlyH^{\zero\frac43}\cdot\frac{1}{\tk}\widetilde{\delta R}^\two(h_{ab}) +\tfrac12\delta\pphisq
						-\curlyH^{\zero\frac43}\cdot\tfrac12\gamma^{ij}\delta\phi_{A,i}\delta\phi_{A,j} \notag\\ 
	&\hspace{0.15\linewidth} + \curlyH^{\zero2}\cdot\tfrac12\delta\phi_A\delta\phi_B\left.\frac{\partial^2V}{\partial\phi_A\partial\phi_B}\right|_{\phibar} \Bigg] \label{Hdens2}
\end{align}
where
\begin{equation} F(T)\equiv \curlyH^{\zero-1}\cdot\left[ \tfrac34\cdot\tk T^2-2V(\phibar)\right]^{-1}
			=\left[2(\tfrac38\cdot\tk T^2-V(\phibar))\pphibar{A}^2\right]^{-\frac12}. \label{FofT}\end{equation}
The details of this procedure are spelled out in appendix \ref{perturbativemethod}. The Hamiltonian determining the perturbative dynamics is
\begin{equation} H_{pert}=\int_{\mathscr{V}}\,d\text{vol}\;\curlyH(x), \label{Hpert}\end{equation}
where $\mathscr{V}$ is the chosen normalisation coordinate volume. (Hence a change in the choice of coordinates on the slice at this stage would imply a change in the limits, leaving value of the Hamiltonian overall unchanged. However, having chosen some coordinates initially one now has that $d$vol$=1\cdot d^3x$ since the notion of scale has been extracted during the Hamiltonian reduction.)
%CLARIFY, E.G. WHERE EXACTLY DO COORDINATES GET FIXED AND WHAT IF WE CHANGED COORDS NOW, E.G. BY A CONSTANT SCALING, HOW WOULD THIS MODIFY $\curlyH$?
%--??-- the coordinates $x$ are necessarily such that the volume element is unity since the notion of scale has been extracted above. IS THIS TRUE? CLARIFY THIS? E.G. WHERE ABOVE EXACTLY DOES THE CHOICE OF COORDINATES GET FIXED IN THIS WAY? OR WHAT IF WE CHANGED COORDS NOW, E.G. BY A CONSTANT SCALING, HOW WOULD THIS MODIFY $\curlyH$?

From this one can obtain the equations of motion,
\begin{align}
 \PD{h_{ab}}{T}  &=\{h_{ab}(x),H_{pert}\} \notag\\
		 &= F(T)\left[2\cdot\tk\nu_{ab}+\curlyH^\zero\cdot\tk Th_{ab}\right], \label{heom}\\
 \PD{\nu^{ab}}{T}&=\{\nu^{ab}(x),H_{pert}\} \notag\\
		 &= F(T)\Bigg[-\curlyH^\zero\cdot\tk T\nu^{ab}  \notag\\
		 &\hspace{0.2\linewidth} -2\curlyH^{\zero\frac43}\cdot\frac{1}{\tk}\left(\delta^{(ab)(cd)}-\tfrac13\gamma^{ab}\gamma^{cd}\right)
			    \gamma^{ij}h_{ic,jd} \label{nueom} \Bigg]
		    %+\curlyH^{\zero\frac43}\cdot\frac{1}{\tk}\int d^3y\;\left(\delta_c^{(a}\delta^{b)}_d-\tfrac13\gamma_{cd}\gamma^{ab}\right)\delta^3(x-y)
			%      \PD{\widetilde{\delta R}^\two}{h_{cd}}(y)\Bigg]. \label{nueom}
\end{align}
where $\delta^{(ab)(cd)}\equiv\delta_m^{(a}\delta^{b)}_n\gamma^{mc}\gamma^{nd}$. The term in the last line is derived from the perturbation of the scale-free curvature, $\de\widetilde{\delta R}^\two/\de h_{ij}$, and is discussed in appendix \ref{curvatureterms}. 

It is worth noting that, since the Hamiltonian density contains only second-order terms, only zeroth-order contributions to the Poisson brackets \ref{hnuPB}, \ref{nunuPB} contribute to first-order terms in equations \ref{heom}, \ref{nueom}. That is, the momentum-momentum Poisson bracket \ref{nunuPB} is effectively canonical, while the position-momentum bracket remains non-canonical due to the `$-\frac13\gamma_{ab}\gamma^{cd}$ term. In other words, for the purposes of first-order perturbation theory the relevant Poisson structure is determined by the background only.

It is easy to confirm using eqs.\ \ref{heom},\ref{nueom} that the tracelessness of $h_{ab}$ and $\nu^{ab}$ is indeed conserved as contraction of eqs.\ \ref{heom} and \ref{nueom} with $\gamma^{ab}$ and $\gamma_{ab}$ respectively reveals. Therefore the constraints $\gamma^{ab}h_{ab}=0$ and $\gamma_{ab}\nu^{ab}=0$ are first class.

The equations for the scalar fields are
\begin{align}
 \PD{\delta\phi_A}{T}(x)   &=\{\delta\phi_A(x),H_{pert}\} = F(T)\cdot\delta\pphi{A}(x) \label{deltaphiEOM}\\
 \PD{\delta\pphi{A}}{T}(x) &=\{\delta\pphi{A}(x),H_{pert}\} \notag\\
			   &= -F(T) \bigg[\curlyH^{\zero \frac43}\int d^3y\; \gamma^{ij}\de_i\left(\delta^3(x-y)\de_j\delta\phi_A(y)\right) \notag\\
			   &\hspace{0.4\linewidth}+\curlyH^{\zero 2}\frac{\de^2V}{\de\phi_A\de\phi_C}\delta\phi_C(x) \bigg] \label{deltapphiEOM}
\end{align}
It is a noteworthy feature that these equations have decoupled from the geometric degrees of freedom. This is a particular feature of the York gauge since matter-field perturbations would otherwise couple with perturbations in the local scale, that is, the metric determinant. Appendix \ref{usefulformulae} discusses what the matter-geometry mixed terms are and how the are eliminated by the York gauge. This result does not generalise to any form of matter, however. In particular, tensor fields coupling to the metric would lead to mixed terms not eliminated by the choice of gauge and therefore a geometric-matter interaction even at the linear level.

By construction equations \ref{heom} and \ref{nueom} (as well as \ref{deltaphiEOM} and \ref{deltapphiEOM}) are linear and therefore may be Fourier analysed. The behaviour of the solutions strongly depends on that of the time-dependent factors $F(T)$ and $\curlyH(T)$, which is in turn dependent on the choice of potential $V(\phi)$. For an analysis of the behaviour of the background, in particular at late times $T\rightarrow0^-$ and for candidates of inflationary potentials, see \citep{Roser2015CosmExtension}. If we consider a free field, explicit functions of time for $F(T)$ and $\curlyH^\zero(T)$ can easily be written down since $\pphi{A}(T)=\pphi{A}(T_0)$ is constant and fixed by the boundary conditions at some time $T_0$:
\begin{align}
F(T) 		 &= \left(\tfrac34\cdot\tk\right)^{-\frac12}|T|^{-1}, \label{FofTfreefield} \\
\curlyH^\zero(T) &= \sqrt{\pphisq(T_0)/(\tfrac34\cdot\tk)}|T|^{-1} = \sqrt{\pphisq(T_0)}\cdot F(T). \label{H_BGfreefield}
\end{align}

For the Fourier analysis we first expand the perturbation variables,
\begin{equation} h_{ab}(x) = \int d^3k\;\xi(k)_{ab}e^{ik\cdot x},\hspace{0.1\linewidth} \nu^{ab}(x) = \int d^3k\;\mu^{ab}(k) e^{ik\cdot x}.\end{equation}
Eq.\ \ref{heom} defines the relationship between $h^\prime_{ab}$ and the momenta $\nu^{cd}$ and, after acting with $\int d^3x\,e^{-il\cdot x}$ on both sides, yields
\begin{equation}\xi_{ab}^\prime (k) = F(T) \cdot\tk\left[ 2\gamma_{ac}\gamma_{bd}\mu^{cd}(k)+\curlyH^\zero T\xi_{ab}(k) \right] \label{xieom}\end{equation}
and is independent of the chosen mode. Eq.\ \ref{nueom} contains the actual dynamics and leads to $k$-dependent terms,
\begin{align} \mu^{ab\prime}(k) 
    &= F(T)\Big[ -\curlyH^\zero\cdot\tk T\mu^{ab}(k) \notag\\ 
    &\hspace{0.15\linewidth} + 2\curlyH^{\zero\frac43}(\tk)^{-1}\left(\delta^{(ab)(cd)}-\tfrac13\gamma^{ab}\gamma^{cd}\right)\gamma^{ij}\xi_{ic}k_jk_d \Big],
\end{align}
Here one can see that for long momentum-space `wavelengths' (small $k$), the first term dominates (although exactly how small $k$ has to be for the first term to dominate depends on the relative sizes of $T$ and $\curlyH^{\zero\frac13}$, that is, the ratio of York time and the linear size of the universe) and the fractional change of each `momentum mode' $\mu^{ab}(k)$ is given by $\curlyH^\zero T$. Since $T<0$ for an expanding universe, and since both $F(T)$ and $\curlyH^\zero$ are positive, the fractional growth of $\mu^{ab}$ is positive. For a theory with free fields, $F(T)\sim-T^{-1}$ and $\mu^{ab}$ just grows with the volume of the universe. On the other hand, for large $k$ the second term dominates and the momentum mode is dynamical.

\section{Quantisation}

Quantisation is the procedure of constructing a new `quantum' theory based on the features of the classical theory, which in turn should form some appropriate limit of the newly constructed quantum theory. However, the general recipe for this construction applies to classical theories with canonical Poisson brackets, a feature the classical cosmological dynamics developed above lacks. In \citep{Roser2015a} we explored the implications of analogous Poisson brackets related to a minisuperspace model for quantisation. The quantisation of general (non-canonical) Poisson brackets is discussed, for example, in \citep{Thiemann2008}.

The primary physical object in a quantum theory is the quantum state, which is represented as a complex function (or functional) defined on the configuration space of the variables of the classical theory and on which operators derived from the classical variables and momenta act, and which evolves according to the Schr\"odinger equation defined by the operator-promoted Hamiltonian. In general the state may also be expressed as a function of other variables, such as the momenta. This is not the case here, as we discussed in section \ref{PertPrinciples}, since the classical momenta do not Poisson commute among themselves, a feature they retain when `promoted' to commutator brackets between operators (eq.\ \ref{Pidontcommute}). Instead due to the asymmetric nature of the $\gtilde_{ab}$ and $\pitilde^{ab}$ only a `position' (that is, field-value) representation can be constructed. If sufficient evidence were found to lend credence to York time as a physically fundamental time, then this asymmetry may be taken to suggest that position is a physically preferred basis, although the conceptual implications do not concern us here.

In generality, promoting eqs.\ \ref{generalgpiPB}, \ref{generalpipiPB} to commutator brackets with dimensionally appropriate factors of $i\hbar$, and the variables $\gtilde_{ab}$ and $\pitilde^{ab}$ to operators $\hat{\gtilde}_{ab}$, $\hat{\pitilde}^{ab}$ respectively leads to the non-canonical operator commutation equations,
\begin{align}
 [\hat\gtilde_{ab}(\vec{x}),\hat\pitilde^{cd}(\vec{y})]  &= i\hbar\left(\delta_a^{(c}\delta^{d)}_b-\tfrac13\hat\gtilde_{ab}\hat\gtilde^{cd}\right)\delta^3(\vec{x}-\vec{y}),
		\label{generalgpiComm} \\
 [\hat\pitilde^{ab}(\vec{x}),\hat\pitilde^{cd}(\vec{y})] &= \frac{i\hbar}3\left(\hat\gtilde^{cd}\hat\pitilde^{ab}-\hat\gtilde^{ab}\hat\pitilde^{cd}\right)\delta^3(\vec{x}-\vec{y}).
		\label{generalpipiComm} 
\end{align}
In principle eq.\ \ref{generalpipiComm} has a factor-ordering ambiguity. However, as long as both `$\hat\gtilde$-$\hat\pitilde$' terms are ordered the same way, the ordering is irrelevant since they can be shown to cancel out using eq.\ \ref{generalgpiComm}. The inverse-metric operator $\hat{\gtilde}^{ab}$ is to be interpreted such that $\hat\gtilde^{ab}\hat\gtilde_{bc}=\delta^a_c$, just like their classical counterparts.

In the position representation the state is given in the form of a functional $\Psi(\gtilde_{ij})$ on the configuration space given by all possible configurations $\gtilde_{ij}(x)$ (here we suppress the dependence of $\Psi$ on the matter variables), which is conformal superspace because of the constraint that the reduced scale $\det(\gtilde_{ij})=1$. The role of the constraint in the quantum theory will be discussed in due course. The defining feature of the position representation is the action of the position operator,
\begin{equation} \hat\gtilde_{ab}\Psi(\gtilde_{ij}) = \gtilde_{ab}\Psi(\gtilde_{ij}). \label{posrepg}\end{equation}
The commutation equations \ref{generalgpiComm}, \ref{generalpipiComm} are then satisfied if the action of the momentum operator $\hat\pitilde^{ab}$ on $\Psi(\gtilde_{ij})$ is
\begin{equation} \hat\pitilde^{ab}\Psi(\gtilde_{ij}) = -i\hbar\big(\delta^{ab}_{cd}-\tfrac13\gtilde^{ab}\gtilde_{ij}\big)\fd{\gtilde_{ij}}. \label{posreppi}\end{equation}

Since the matter variables are canonical and furthermore commute with the geometric variables, their representation is not restricted to a particular basis. While one could imagine `mixed' bases such that $\Psi=\Psi(\gtilde_{ab},\pphi{A})$, here we assume the full position basis where $\Psi=\Psi(\gtilde_{ab},\phi_A)$, although since our focus in on the geometric degrees of freedom, this is not important for our discussion here.

From this general framework one obtains the representation of the perturbation operator-promoted variables $\hat h_{ab}$ and $\hat\nu^{ab}$ for a wave functional in the position basis,
\begin{align}
 \hat h_{ab}\Psi(h_{ij})  &= h_{ab}\Psi(h_{ij}) \label{posreph} \\
 \hat\nu^{ab}\Psi(h_{ij}) &= -i\hbar\big(\delta^{ab}_{cd}-\tfrac13(\gamma^{ab}-\gamma^{am}h_{mn}\gamma^{nd})(\gamma_{cd}+h_{cd})\big)\fd{h_{cd}} \Psi(h_{ij}) \notag\\
			  &= -i\hbar\big(\delta^{ab}_{cd}-\tfrac13\gamma^{ab}\gamma_{cd}-\tfrac13\gamma^{ab}h_{cd}+\tfrac13h^{ab}\gamma_{cd} +\text{h.o.}\big)
			  \fd{h_{cd}}\Psi(h_{ij}).    \label{posrepnu}
\end{align}

The quantum dynamics is determined by the operator form of the Hamiltonian given by \ref{Hdens2}-\ref{Hpert}. However, the Hamiltonian involves explicit functions of time obtained from solutions of the classical dynamics of the background. For the purposes of the quantum theory one could choose to adopt the same form, contending with quantised perturbations on a classical background. Here we are interested in a cosmological theory that is fully quantum and we therefore also quantise the background. Standard quantum mechanics does not, however, provide solutions that could be substituted for the classical ones since there are no trajectories in configuration space, as the object of interest is the state or wavefunction. This conundrum may be resolved by employing the trajectories obtained in de~Broglie-Bohm (or `Pilot-Wave') theory \citep{deBroglie1928,Bohm1952,Holland1993}. De~Broglie-Bohm trajectories have previously been employed in cosmological perturbation theory \citep[e.g.][]{PeterPinhoPintoneto2007,PeterPintoNeto2008}. While non-equilibrium de~Broglie-Bohm theory when applied to cosmological applications may lead to a distinct phenomenology \citep{Valentini2010InflCosm,ValentiniColin2015a,Valentini2015a, UnderwoodValentini2015} it agrees perfectly in its statistical predicitons with standard quantum mechanics in its equilibrium form and can therefore be employed as a mathematical tool even if one is unwilling to make any form of ontological commitment. The fact that de~Broglie-Bohm regards the position basis as privileged is furthermore consistent with its special status in the present formalism.

We fully solved the background dynamics for a single scalar field in \citep{RoserValentini2014a}, where we found that the quantum trajectories do, in fact, match the classical ones. Furthermore, the Hamiltonian expectation values --- physically the volume of the universe up to a constant, analogous to the notion of energy in quantum mechanics of more conventional systems --- also match the evolution of the classical volume. A similar quantum-classical correspondence was established using de~Broglie-Bohm for other appropriate matter content with standard cosmological time and outside the reduced-Hamiltonian picture in \citep{John2015}. While such results are encouraging, the equality of classical and quantum trajectories is not necessary for the method described here to be applied. However, the fact that the two do match for certain scenarios may explain why quantising only the perturbations and leaving the background classical, as is often done, is empirically successful. For the case of a free field we found $\phi(T)\sim\ln|T|$ and $\oppr{\Psi}{\hat H_{BG}}{\Psi}\sim |T|^{-1}$ with the factors of proportionality depending on the chosen initial state \citep{RoserValentini2014a}.

Having solved the background one has explicit functions of time $\phibar_A(T)$ and $\pphibar{A}(T)$, and the quantum analogue of $\curlyH^\zero(T)$ is given by the expectation value of the background Hamiltonian. This fully defines the perturbation Hamiltonian in the quantum theory.

The dynamics is further defined by the presence of the constraints implied by the York gauge, that is, the choice of gauge such that the foliation is parameterised by York time exactly rather than merely at zeroth order. Classically this condition implied that $h_{ab}$ and $\nu^{ab}$ are traceless, which according to the Dirac quantisation procedure leads to constraints on the set of physical states $\Psi_{phys}$,
\begin{align}
 \gamma^{ab}\hat h_{ab} \Psi_{phys} &= 0 \label{qhconstraint} \\
 \gamma_{ab}\hat\nu^{ab}\Psi_{phys} &= 0. \label{qnuconstraint}
\end{align}
In the position representation the first of these implies that $\Psi_{phys}$ must vanish off the classical constraint surface defined by $h_{ij}=0$, reducing the dimensionality of the physical configuration space by one per spatial point.\footnote{The true number of physical degrees of freedom is, of course, further reduced by the 3-conformal-diffeomorphism invariance.} From the linearity of the Schr\"odinger equation it follows that the vanishing of $\Psi_{phys}$ off the surface is consistent with its time evolution. 

The physical wave functional is therefore necessarily discontinuous in the full (unconstrained) configuration space. This is nonetheless not in conflict with the action of the momenta $\hat\nu^{ab}$ as derivatives since they act tangent to the constraint surface, just as their classical generator counterparts. In \citep{Roser2015a} we discussed this feature in some detail with an appropriate change of variables for a minisuperspace model.

The momentum-trace constraint \ref{qnuconstraint} does not restrict the set of possible physical states any further since it is identically satisfied by the choice of representation of $\hat\pitilde^{ab}$ even in the general case \ref{posreppi}.

The momentum operator is Hermitian only when considered to zeroth order (meaning the first two terms in \ref{posrepnu}), where its form is fully determined by the background. At this order a momentum representation exists, although note that the representation of the momentum in the position basis is still not the canonical one. The classical analogue is that the Poisson brackets only contribute to the linear perturbation equations at zeroth order as discussed above. The Hamiltonian may be made Hermitian by choosing the symmetric ordering for the mixed term,
\begin{equation} F(T)\curlyH^\zero\cdot\tk T\;\cdot\tfrac12(\hat h_{ij}\hat\nu^{ij} + \hat\nu^{ij}\hat h_{ij}).\end{equation} 

The reason hermiticity cannot be established at all orders is in part that in the derivation of the Hamiltonian it was assumed that the perturbation variables $h_{ab}$ and $\nu^{ab}$ are sufficiently small for a second-order Hamiltonian to adequately describe the dynamics. The statement that the momentum or Hamiltonian operators be Hermitian however is, algebraically speaking, a claim involving the functional integral over all allowed functions $h_{ij}$, including those where individual components may be large. Therefore one would have to apply finite (and adequately small) limits to the functional integral, or use an appropriate attenuation functional, representing the notion that physical wavefunctionals must become small for functions $h_{ij}(x)$ with large upper bounds. However, even then one can at best hope to establish Hermiticity approximately. This issue does not depend on whether or not one restricts the functional integration to the constraint surface, whereas in the non-perturbative finite-dimensional model developed in \citep{Roser2015a} hermiticity can be established if one does make that restriction.

In practice one might ignore these issues and consider expression \ref{posrepnu} to leading order only, so that the Hamiltonian is Hermitian and a well-defined probability current exists. %Some of the implications of dealing with non-Hermitian operators are discussed in the conclusion \ref{conclusion}. 
One consequence of this is however that the geometric momentum operator $\hat\nu^{ab}$ is therefore only a quantum observable at leading order. At this order, however, the quantum theory is well defined.

%DISCUSS HERMITICITY OF OPERATORS\\
%-- $\nu$ Hermitian at leading order on full space, hence so is $\hat{\curlyH}^\two$ provided mixed term has symmetric factor ordering \\
%-- $\nu$ is not Hermitian on full config space at higher order, but possibly is over the constraint surface (how to show this to be true/false? Integration over a function space is hard ... some clever change of variables similar to other paper?)\\
%-- If momenta not Hermitian at all order, can a ordering in the Hamiltonian be found, so it is? And then momenta would not be `observables' ---what does this mean physically speaking? Not measurable? We only measure positions ever anyway.  If the Hamiltonian is not Hermitian at higher orders, would we have trouble with a conserved probability?
%-- Just ignore foundational issues and work with leading-order Hermiticity?

%EIGENFUNCTIONS --- probably not easily solvable since field equation, so don't follow this path

%...[[IN UPDATED VERSION: commutator brackets for $\hat\xi_{ab}$ and $\hat\mu^{cd}$ (like in standard QFT) --- solutions? Trajectories for perturbations?]]...

\section{Conclusion}\label{conclusion}

General relativity itself does not have a physically privileged time parameter. However, there may be reasons to consider the possibility that an underlying such parameter exists and that it is merely hidden by the fact that the theory's laws \emph{may} be expressed in a four-covariant manner. Yet the four-covariance, in particular the refoliation and reparameterisation invariance leads to problems during attempted quantisation. One such problem follows from the vanishing of the Hamiltonian constraint, a direct consequence of the time-reparameterisation invariance. This may be overcome by making an informed guess what the underlying privileged time parameter may be and performing a Hamiltonian reduction in order to arrive at a physically meaningful, non-vanishing Hamiltonian density. Different choices of time will lead to phenomenologically identical classical theories. However, since time is treated differently from the other physical variables during quantisation, different quantum theories with possibly different phenomenology may result. It may therefore be possible, at least in principle in some cases, to discern between theories with different time parameters empirically.

York time is a viable candidate for a physically privileged time because of its special role in the initial-value problem as well as other reasons. Here we solved the Hamiltonian constraint for the York-time associated physical Hamiltonian in the context of cosmological perturbation theory up to second order, leading to linear equations of motion. However, since the Poisson brackets of the reduced degrees of freedom are not canonical (in particular, the momentum-momentum bracket does not vanish), the quantum theory does not have a momentum representation. Hence, if York time is indeed a physically privileged time, the `position' basis (the basis given by the set of possible three-metric-field configurations on space) is physically more fundamental, opening up a variety of conceptual questions for quantum theory, which we will not discuss here. The physically privileged position basis is however consistent with the approach of the de~Broglie-Bohm formulation of quantum theory, which we employed here as a mathematical tool allowing us to quantise the cosmological background as well as the perturbative degrees of freedom.

The position-space representation of the momentum operators differs from its conventional form due to the modified Poisson (and hence commutator) brackets. The momenta are not Hermitian, which implies that they are not quantum `observables'. This however is not a problem for the physical viability of the quantum theory. As pointed out by Feynman and Hibbs in the context of the path-integral formulation \citep[ch.\ 5]{FeynmanHibbs1965} and as often emphasised by Bell in other contexts \citep{Bell1987}, all actual physical measurements are ultimate measurements of position, such as position after some time interval in time-of-flight measurements or the position of pointers in our equipment. If we consider it essential that the Hamiltonian be Hermitian\footnote{Hermiticity of the Hamiltonian ensures, of course, a variety of properties such as a well-defined probability current. However, a modification of the choice of inner product can still lead to a well-defined notion of probability even for non-Hermitian Hamiltonians. An example of this is the theory of $PT$-symmetric Hamiltonians, where the inner product is defined dynamically and depends on the Hamiltonian \citep{BenderIntro2005}.}%However, in the de~Broglie-Bohm approach ultimately only the well-definedness of the trajectories matters. Furthermore, there is only one universe, so the meaning of probability in this context is somewhat unclear anyway. Not being able to describe the evolution of an ensemble via a probability density may therefore not be a fatal flaw of the theory.} 
then the non-hermiticity of the momenta implies that momentum measurements cannot be described by a conventional von-Neumann measurement Hamiltonian, but this does not preclude the existence of other, Hermitian interaction Hamiltonians allowing us experimental access to the momentum-like quantities. 

%We must, of course, remember that `quantisation' is only a recipe to arrive at a theory with quantum properties from a theory with classical properties such that the classical theory forms an appropriate limit of the quantum theory in certain regimes. The quantisation method used here is, while appearing plausible given the success of following such recipes in other cases, therefore ultimately only a guess, a guess that may be incorrect even if York time does have special physical status.

This paper has been concerned with formulating the classical and quantum perturbation theory. The next step must now be a closer analysis of the dynamics, in particular at the quantum level, in order to make connection to the predictions of conventional cosmological perturbation theory and empirical data where possible.
%TO DO\\
%-- Recap procedure\\
%-- Importance of trajectories $\rightarrow$ two separate hints for foundational importance / privileged status of `position' space (i.e.\ conformal superspace) over momentum space etc.\\
%-- Choice of time matters for physical content (and phenomenology) of quantum theory\\
%-- HERE (preliminary version): Want further dynamical analysis, predictions for certain regimes [[LATER (updated version): solutions, certain limits, relation to standard pert theory]]

\section*{Acknowledgements}
The author would like to thank Antony Valentini for multiple helpful discussions and a review of this manuscript.

%-------------------------------------------------
\appendix
\section{Expansion of relevant terms} \label{usefulformulae}
Prior to application of the York gauge (that is, if York time is only implemented at the background level and one retains the full gauge freedom of standard perturbation theory), expressions \ref{Rterms}-\ref{Vterms} are
\begin{align}
 [\lqu R\rqu] &=-\frac1\tk\left(\tilde{\Rbar}+\widetilde{\delta R}^\one +\eta^\one\tilde{\Rbar}+\widetilde{\delta R}^\two +\eta^\one\widetilde{\delta R}^\one +\eta^\two\tilde{\Rbar} \right) \\
 [\lqu\nabla\phi\rqu] &= \frac12\gamma^{ij}\delta\phi_{A,i}\delta\phi_{A,j} \\
 [\lqu p_\phi\rqu] &= \frac12\Big(\pphibar{A}^2-\eta^\one\pphibar{A}^2+2\pphibar{A}\delta\pphi{A} \notag\\
		    &\hspace{0.2\linewidth}+\big(-\eta^\two+(\eta^\one)^2\big)\pphibar{A}^2-2\eta^\one\pphibar{A}\delta\pphi{A}+\delta\pphi{A}^2\Big) \label{pphitermsgeneral}\\
 [\lqu V\rqu] &= V(\phibar)+\eta^\one V(\phibar)+\delta\phi_A\left.\PD{V}{\phi_A}\right|_{\phibar} \notag\\ 
		    &\hspace{0.2\linewidth}+\eta^\two V(\phibar) +\eta^\one\delta\phi_A\left.\PD{V}{\phi_A}\right|_{\phibar}
			  +\frac12\delta\phi_A\delta\phi_B\left.\frac{\partial^2V}{\partial\phi_A\partial\phi_B}\right|_{\phibar}, \label{Vtermsgeneral}
\end{align}
where
\begin{align} \eta^\one &= \tfrac12 h_{ab}\gamma^{ab} \\
	      \eta^\two &= \tfrac18(h_{ab}\gamma^{ab})^2 - \tfrac14 h_{ac}h_{bd}\gamma^{ab}\gamma^{cd}
\end{align}
are the first and second order fractional perturbation in the metric. In the York gauge these are set to zero. Note that eqs.\ \ref{pphitermsgeneral} and \ref{Vtermsgeneral} contain mixed terms, which would lead to a coupling between matter and geometric perturbation in the linearised equations of motion. However, in the York gauge the perturbation is set to zero, eliminating exactly those terms.

The expressions for the perturbative expansion of $\tilde{R}$ are
\begin{align}
 \tilde{\Rbar} &= \gbar^\frac13\Rbar=\gbar^\frac13\gbar^{ij}\Rbar_{ij} = \gamma^{ij}R_{ij}, \\
 \widetilde{\delta R}^\one &= \gbar^\frac13\delta R^\one = \gbar^\frac13\delta(g^{ij}R_{ij})^\one = \gamma^{ij}\delta R^\one_{ij} + (-\gamma^{ik}h_{kl}\gamma^{lj})\Rbar_{ij}, \\
 \widetilde{\delta R}^\two &= \gbar^\frac13\delta R^\two = \gbar^\frac13\delta(g^{ij}R_{ij})^\two \notag\\
		       &= \gamma^{ij}\delta R^\two_{ij} + (-\gamma^{ik}h_{kl}\gamma^{lj})\delta R^\one_{ij}   + (-\gamma^{ik}h_{kl}\gamma^{lm}h_{mn}\gamma^{nj})\Rbar_{ij}.
\end{align}
with
\begin{align}
  \delta R^\one_{ij} &= \de_k\delta\Gamma^{\one k}_{ij}-\de_i\delta\Gamma^{\one k}_{jk} \\
  \delta R^\two_{ij} &=	  \de_k\delta\Gamma^{\two k}_{ij}-\de_i\delta\Gamma^{\two k}_{jk}
		+\delta\Gamma^{\one k}_{ij}\delta\Gamma^{\one l}_{kl}-\delta\Gamma^{\one k}_{il}\delta\Gamma^{\one l}_{jk},
\end{align}
where
\begin{align}
 \delta\Gamma^{\one q}_{rs} &= \tfrac12\gamma^{qt} (h_{tr,s}+h_{ts,r}-h_{rs,t} ) \\
 \delta\Gamma^{\two q}_{rs} &= -\gamma^{qu}h_{uv}\delta\Gamma^{\one v}_{rs}
\end{align}
denote the perturbation in the Levi-Civita connection.

\section{Perturbative approach to solving the Hamiltonian constraint to second order} \label{perturbativemethod}
Writing $x\equiv\curlyH^\frac13$, eq.\ \ref{HCEq}, after multiplication by $\curlyH=x^3$, has the form
\begin{equation} 0 = ax^6 + bx^4 + cx^3 + d, \label{xequation}\end{equation}
where, separating different orders visually,
\begin{alignat}{4}
 &a= &&-\tfrac38\cdot\tk T^2+ V(\phibar)  \quad&& +\delta\phi_A\left.\PD{V}{\phi_A}\right|_{\phibar} 
		  \quad&& +\tfrac12\delta\phi_A\delta\phi_B\left.\frac{\partial^2V}{\partial\phi_A\partial\phi_B}\right|_{\phibar} \\
 &b= && &&-\frac{1}{\tk}\widetilde{\delta R}^\one \quad&&-\frac{1}{\tk}\widetilde{\delta R}^\two +\tfrac12\gamma^{ab}\delta\phi_{A,a}\delta\phi_{A,b} \\
 &c= && && && \tk T h_{ab}\nu^{ab} \\
 &d= &&  \tfrac12\pphibar{A}^2 \quad&& +\pphibar{A}\delta\pphi{A} \quad&& +\delta\pphi{A}^2 + \tk\gamma_{ac}\gamma_{bd}\nu^{ab}\nu^{cd}.
\end{alignat}
Here we assumed a spatially flat background, so that $\tilde{\Rbar}=0$. As in the text we let $a^\zero$ denote the zeroth order term of $a$ and so on. Similarly, we expand the sought function order by order, $x=x^\zero+x^\one+x^\two$. 

One first solves the zeroth-order equation,
\begin{equation} 0 = a^\zero x^{\zero6}+d^\zero,\end{equation}
so that 
\begin{equation} x^{\zero6} = -\frac{d^\zero}{a^\zero} = \frac{\frac12\bar{p}_\phi^2}{\frac38\cdot\tk T^2-V(\phibar)}, \label{x0solution}\end{equation}
consistent with the result in \citep{RoserValentini2014a}. The first order equation is trivial since $a^\one$, $b^\one$ and $d^\one$ vanish when the background equations are satisfied, hence $x^\one=0$. Then the second-order equation is
\begin{equation} 0=a^\two x^{\zero6}+6a^\zero x^{\zero5} x^\two+b^\two x^{\zero4}+c^\two x^{\zero3}+d^\two, \end{equation}
an equation linear in $x^\two$, giving after substitution of the coefficients and the background solution \ref{x0solution},
\begin{align}
 x^\two &=  \Bigg[ \frac12\delta\phi_A\delta\phi_B\left.\frac{\partial^2V}{\partial\phi_A\partial\phi_B}\right|_{\phibar}x_0^6
		    -\left(\frac{1}{\tk}\widetilde{\delta R}^\two+\frac12\gamma^{ij}\delta\phi_{A,i}\delta\phi_{A,j}\right)x_0^4  \notag\\
	&\hspace{0.05\linewidth} +\tk Th_{ij}\nu^{ij}x_0^3+\delta\pphi{A}^2+\tk\gamma_{ik}\gamma_{jl}\nu^{ij}\nu^{kl} \Bigg]
				\Big[\left(\tfrac38\tk T^2-V(\phibar)\right)x_0^5\Big]^{-1}.
\end{align}
The Hamiltonian density at second order is then
\begin{equation} \curlyH=(x^\zero+x^\two)^3 = x^{\zero3} + 3x^{\zero2}x^\two. \label{fullHsolution}\end{equation}

\section{Terms relating to perturbation of the curvature}\label{curvatureterms}

Derivation of the classical equations of motion involves the term 
\begin{align}\{\nu^{ab}(y),\int d^3x\,\widetilde{\delta R}^\two(x)\} 
	  &= \int d^3x\;\{\nu^{ab}(y),\widetilde{\delta R}(x)\} \notag\\
	  &= \int d^3x\; \{\nu^{ab}(y),h_{cd}(x)\}\;\PD{\widetilde{\delta R}^\two}{h_{cd}}. \label{PBnudeltaR2}
\end{align}
For the flat background assumed in the text and expressed in frame with Cartesian coordinates, one has
\begin{align} \widetilde{\delta R}^\two 
  &= \gamma^{ij}\delta R_{ij} + (-\gamma^{ik}h_{kl}\gamma^{lj})\delta R^\one_{ij} \notag\\
  &= \gamma^{ij}\big(\de_k\delta\Gamma^{\two k}_{ij} - \de_i\delta\Gamma^{\two k}_{jk} 
	      + \delta\Gamma^{\one k}_{ij}\delta\Gamma^{\one l}_{kl} -\delta\Gamma^{\one k}_{il}\delta\Gamma^{\one l}_{jk} \big) \notag\\
  &\hspace{0.05\linewidth} -\gamma^{ik}h_{kl}\gamma^{lj}\big(\de_k\delta\Gamma^{\one k}_{ij}-\de_i\delta\Gamma^{\one k}_{jk} \big).
\end{align}

Since $\widetilde{\delta R}^\two$ is first order, the Poisson bracket only contributes at zeroth order and is therefore spatially constant and may be taken outside the integral. The remaining term is then, abbreviating $\delta^{(3)}\equiv \delta^{(3)}(x-y)$,
\begin{align} &\int d^3x\;\delta^{(3)}\PD{\widetilde{\delta R}^\two}{h_{cd}}(y) \notag\\
 &\quad= \int d^3y\;\Bigg[ \tau^{\one k.cd}_{ij.k}[\delta^{(3)}]\gamma^{ij} - \tau^{one k.cd}_{jk.i}[\delta^{(3)}]\gamma^{ij} 
	    + \sigma^{\zero k.cd}_{ij}[\delta^{(3)}\delta\Gamma^{\one l}_{kl}] \notag \\
	&\hspace{0.2\linewidth} +\sigma^{\zero l.cd}_{kl}[\delta^{(3)}\delta\Gamma^{\one k}_{ij}] -\sigma^{\zero k.cd}_{il}[\delta^{(3)}\delta\Gamma^{\one l}_{jk}]\gamma^{ij} \notag\\
	&\hspace{0.2\linewidth}	-\sigma^{\zero l.cd}_{jk}[\delta^{(3)}\delta\Gamma^{\one k}_{il}]\gamma^{ij} 
				- \gamma^{im}\delta^{cd}_{mn}\gamma^{nj}(\de_k\delta\Gamma^{\one k}_{ij}-\de_i\delta\Gamma^{\one k}_{jk}) \notag\\
	&\hspace{0.2\linewidth} -\tau^{\zero k.cd}_{ij.k}[\delta^{(3)}h^{ij}] + \tau^{\zero k.cd}_{jk.i}[\delta^{(3)}h^{ij}]\Bigg] \label{dR2dh}
\end{align}
with
\begin{align}
 \sigma^{\zero q.ab}_{rs}[f] &= -\tfrac12\gamma^{qt} (\delta^{ab}_{tr}\de_s + \delta^{ab}_{ts}\de_r - \delta^{ab}_{rs}\de_t ) f \\
 \sigma^{\one q.ab}_{rs}[f] &= -\gamma^{q(a}\delta\Gamma^{\one v)}_{rs} \cdot f 
				    + \tfrac12\gamma^{vt}\gamma^{qu}(\delta^{ab}_{tr}\de_s + \delta^{ab}_{ts}\de_r - \delta^{ab}_{rs}\de_t ) (h_{uv}f) \\
 \tau^{\zero q.ab}_{rs.i}[f] &=  \tfrac12\gamma^{qt}(\delta^{ab}_{tr}\de_i\de_s + \delta^{ab}_{ts}\de_i\de_r-\delta^{ab}_{rs}\de_i\de_t ) f \\
 \tau^{\one q.ab}_{rs.i}[f] &= \delta^{ab}_{uv}\gamma^{qu}\de_i(\delta\Gamma^{\one v}_{rs} f) - \tfrac12\delta^{ab}_{uv}\gamma^{qu}\gamma^{vt}(h_{tr,si}+h_{ts,ri}-h_{rs,ti})\cdot f \notag\\
			    &\hspace{0.05\linewidth} +\tfrac12\gamma^{qu}\gamma^{vt} (\delta^{ab}_{tr}\de_i\de_s + \delta^{ab}_{ts}\de_i\de_r - \delta^{ab}_{rs}\de_i\de_t ) (h_{uv} f),
\end{align}
defined as
\begin{align}
 \int d^3x\;\sigma^{\zero q.ab}_{rs}[f] &= \int d^3x\;\PD{\Gamma^{\one q}_{rs}}{h_{ab}} f \\  
 \int d^3x\;\sigma^{\one q.ab}_{rs}[f] &= \int d^3x\;\PD{\Gamma^{\two q}_{rs}}{h_{ab}} f \\
 \int d^3x\;\tau^{\zero q.ab}_{rs.i}[f] &= \int d^3x\;\pd{h_{ab}} \left[\de_i\Gamma^{\one q}_{rs} \right] f \\  
 \int d^3x\;\tau^{\one q.ab}_{rs}[f] &= \int d^3x\;\pd{h_{ab}} \left[\de_i\Gamma^{\two q}_{rs} \right] f,
\end{align}
where $f$ is any appropriate functions of the spatial coordinates. The mismatch of notational superscripts $(n)$ was chosen in order to remain consistent with their role of identifying the order of perturbation, which is reduced by one as a result of the differentiation with respect to $h_{ab}$.

Expression \ref{dR2dh} contains a number of total derivatives as well as a significant degree of symmetry and evaluates to the simpler expression
\begin{equation} \int d^3x\;\delta^{(3)}\PD{\widetilde{\delta R}^\two}{h_{cd}}(y) = -2\delta^{(cd)(kt)}\gamma^{ij}h_{it,jk} \end{equation}
after all boundary terms have been dropped, which follows from assuming appropriate boundary conditions, and the constraint $\gamma^{ij}h_{ij}(x)=0$ has been used.

%_________________________________________________________________________________
%BIBLIOGRAPHY

\bibliographystyle{unsrtnat}	%IF STYLE CHANGED, FIRST DELETE .AUX AND .BBL FILES BEFORE RECOMPILING
\bibliography{../../Bibloi}	%ENTER CORRECT RELATIVE PATH TO BIB-FILE

\begin{thebibliography}{45}
\providecommand{\natexlab}[1]{#1}
\providecommand{\url}[1]{\texttt{#1}}
\expandafter\ifx\csname urlstyle\endcsname\relax
  \providecommand{\doi}[1]{doi: #1}\else
  \providecommand{\doi}{doi: \begingroup \urlstyle{rm}\Url}\fi

\bibitem[Kucha\v{r}(2011)]{Kuchar2011}
Karel Kucha\v{r}.
\newblock {Time and Interpretations of Quantum Gravity}.
\newblock \emph{International Journal of Modern Physics D}, 20:\penalty0 3--86,
  2011.

\bibitem[Anderson(2012)]{Anderson2012}
E.~Anderson.
\newblock {The problem of time in quantum gravity}.
\newblock In V.~R. Frignanni, editor, \emph{{Classical and Quantum Gravity:
  Theory, Analysis and Applications}}. Nova (New York), 2012.
\newblock gr-qc: 1206.2403.

\bibitem[Arnowitt et~al.(1962)Arnowitt, Deser, and Misner]{ADM1962}
R.~Arnowitt, S.~Deser, and C.~W. Misner.
\newblock The dynamics of general relativity.
\newblock In L.~Witten, editor, \emph{Gravitation: an introduction to current
  research}. Wiley, 1962.

\bibitem[DeWitt(1967)]{DeWitt1967}
Bryce DeWitt.
\newblock {Quantum Theory of Gravity I: The Canonical Theory}.
\newblock \emph{Physical Review}, 160\penalty0 (5):\penalty0 1113, 1967.

\bibitem[Unruh and Wald(1989)]{UnruhWald1989}
William~G. Unruh and Robert~D. Wald.
\newblock Time and the interpretation of canonical quantum gravity.
\newblock \emph{Physical Review D}, 40\penalty0 (8):\penalty0 2598, October
  1989.

\bibitem[Rovelli(2007)]{Rovelli2007}
Carlo Rovelli.
\newblock \emph{Quantum Gravity}.
\newblock Cambridge University Press, 2007.

\bibitem[Brown and Kucha\v{r}(1995)]{BrownKuchar1995}
J.\ Brown and Karel Kucha\v{r}.
\newblock Dust as a standard of space and time in canonical quantum gravity.
\newblock \emph{Physical Review D}, 51:\penalty0 5600--5629, 1995.

\bibitem[Husain and Pawlowski(2012)]{HusainPawlowski2012}
Viqar Husain and Tomasz Pawlowski.
\newblock {Time and a physical Hamiltonian for quantum gravity}.
\newblock \emph{Physical Review Letters}, 108, 2012.

\bibitem[York(1972)]{York1972}
James York.
\newblock Role of conformal three-geometry in the dynamics of gravitation.
\newblock \emph{Physical Review Letters}, 28:\penalty0 1082--1085, 1972.

\bibitem[Choquet-Bruhat and York(1980)]{ChoquetBruhatYork1980}
Yvonne Choquet-Bruhat and James York.
\newblock {The Cauchy Problem}.
\newblock In Achim Held, editor, \emph{General Relativity and Gravitation I}.
  Plenum, 1980.

\bibitem[York(1971)]{York1971}
James York.
\newblock Gravitational degrees of freedom and the initial-value problem.
\newblock \emph{Physical Review Letters}, 26:\penalty0 1656--1658, 1971.

\bibitem[York(1973)]{York1973}
James York.
\newblock Conformally invariant orthogonal decomposition of symmetric tensors
  on riemannian manifolds and the initial-value problem of general relativity.
\newblock \emph{Journal of Mathematical Physics}, 14:\penalty0 456--464, 1973.

\bibitem[Barbour et~al.(2002)Barbour, Foster, and {\'O
  Murchadha}]{BarbourFosterOMurchadha2002}
J.~Barbour, B.~Foster, and N.~{\'O Murchadha}.
\newblock {Relativity without Relativity}.
\newblock \emph{Classical and Quantum Gravity}, 19:\penalty0 3217--3248, 2002.
\newblock gr-qc: 0012089.

\bibitem[Anderson et~al.(2003)Anderson, Barbour, Foster, and {\'O
  Murchadha}]{AndersonBarbourFosterOMurchadha2003}
Edward Anderson, Julian Barbour, Brendan Foster, and Niall {\'O Murchadha}.
\newblock {Scale-invariant gravity: Geometrodynamics}.
\newblock \emph{Classical and Quantum Gravity}, 20:\penalty0 1571, 2003.
\newblock gr-qc: 0211022.

\bibitem[Anderson et~al.(2005)Anderson, Barbour, Foster, Kelleher, and {\'O
  Murchadha}]{AndersonBarbourFosterKelleherOMurchadha2005}
Edward Anderson, Julian Barbour, Brendan Foster, Bryan Kelleher, and Niall {\'O
  Murchadha}.
\newblock {The physical gravitational degrees of freedom}.
\newblock \emph{Classical and quantum gravity}, 22:\penalty0 1795--1802, 2005.
\newblock gr-qc: 0407104.

\bibitem[Gomes et~al.(2011)Gomes, Gryb, and Koslowski]{GomesGrybKoslowski2011}
Henrique Gomes, Sean Gryb, and Tim Koslowski.
\newblock Einstein gravity as a 3d conformally invariant theory.
\newblock \emph{Classical and Quantum Gravity}, 28:\penalty0 045004, 2011.
\newblock gr-qc: 1010.2481.

\bibitem[Barbour et~al.(2013)Barbour, Koslowski, and
  Mercati]{BarbourKoslowskiMercati2013}
Julian Barbour, Tim Koslowski, and Flavio Mercati.
\newblock {The gravitational origin of the arrow of time}.
\newblock 2013.
\newblock gr-qc: 1310.5167.

\bibitem[Mercati(2014)]{Mercati2014}
Flavio Mercati.
\newblock A shape dynamics tutorial.
\newblock 2014.
\newblock gr-qc: 1409.0105v1.

\bibitem[Gomes(2014)]{Gomes2014}
Henrique Gomes.
\newblock {A Birkhoff Theorem for Shape Dynamics}.
\newblock \emph{Classical and Quantum Gravity}, 31:\penalty0 085008, 2014.
\newblock gr-qc: 1305.0310.

\bibitem[Gomes and Herczeg(2014)]{GomesHerczeg2014}
Henrique Gomes and Gabriel Herczeg.
\newblock {A Rotating Black Hole Solution for Shape Dynamics}.
\newblock \emph{Classical and Quantum Gravity}, 31:\penalty0 175014, 2014.
\newblock gr-qc: 1310.6095v4.

\bibitem[Valentini(1996)]{Valentini1996}
Antony Valentini.
\newblock Pilot-wave theory of fields, gravitation and cosmology.
\newblock In J.~T. Cushing, A.~Fine, and S.~Goldstein, editors, \emph{Bohmian
  mechanics and quantum theory: an appraisal}. Kluwer, 1996.

\bibitem[Qadir and Wheeler(1985)]{QadirWheeler1985}
A.~Qadir and J.A. Wheeler.
\newblock In \emph{{From $SU(3)$ to Gravity}}. Cambridge University Press,
  1985.

\bibitem[Roser and Valentini(2014)]{RoserValentini2014a}
Philipp Roser and Antony Valentini.
\newblock {Classical and quantum cosmology with York time}.
\newblock \emph{Classical and Quantum Gravity}, 31:\penalty0 245001, 2014.
\newblock gr-qc: 1406.2036.

\bibitem[Roser(2015{\natexlab{a}})]{Roser2015a}
Philipp Roser.
\newblock {Quantum mechanics on York slices}.
\newblock 2015{\natexlab{a}}.
\newblock qr-qc: 1507.01556.

\bibitem[Muhkanov et~al.(1992)Muhkanov, Feldman, and
  Brandenberger]{MukhanovFeldmanBrandenberger1992}
V.F. Muhkanov, H.A. Feldman, and R.H. Brandenberger.
\newblock {Theory of cosmological perturbations}.
\newblock \emph{Physics Reports}, 215:\penalty0 203, 1992.

\bibitem[Mukhanov(2005)]{Mukhanov2005}
V.~Mukhanov.
\newblock \emph{The Physical Foundations of Cosmology}.
\newblock Cambridge University Press, 2005.

\bibitem[Bardeen(1980)]{Bardeen1980}
James Bardeen.
\newblock {Gauge-invariant cosmological perturbations}.
\newblock \emph{Physical Review D}, 22\penalty0 (8):\penalty0 1882--1905, 1980.

\bibitem[de~Broglie(1928)]{deBroglie1928}
Louis de~Broglie.
\newblock In \emph{{\'Electrons et Photons: Rapports et Discussions de
  Cinqui\`eme Conseil de Physique}}. Gauthier-Villars (Paris), 1928.
\newblock {English translation in G.\ Bacciagaluppi and A.\ Valentini,
  \emph{Quantum Theory at the Crossroads: Reconsidering the 1927 Solvay
  Conference}, Cambridge University Press, 2009, quant-ph: 0609184}.

\bibitem[Bohm(1952)]{Bohm1952}
David Bohm.
\newblock {A Suggested Interpretation of the Quantum Theory in Terms of
  `Hidden' Variables. I and II}.
\newblock \emph{Physical Review}, 85:\penalty0 166--179, 180--194, 1952.

\bibitem[Holland(1993)]{Holland1993}
Peter~R. Holland.
\newblock \emph{{The Quantum Theory of Motion: an Account of the
  de~Broglie-Bohm Causal Interpretation of Quantum Mechanics}}.
\newblock Cambridge University Press, 1993.

\bibitem[Pinto-Neto et~al.(2012)Pinto-Neto, Santos, and
  Struyve]{Pinto-NetoSantosStruyve2012}
Nelson Pinto-Neto, Grasiele Santos, and Ward Struyve.
\newblock {Quantum-to-classical transition of primordial cosmological
  perturbations in de~Broglie-Bohm quantum theory}.
\newblock \emph{Physical Review D}, 85:\penalty0 082506, 2012.
\newblock gr-qc: 1110.1339.

\bibitem[Pinto-Neto et~al.(2014)Pinto-Neto, Santos, and
  Struyve]{Pinto-NetoSantosStruyve2014}
Nelson Pinto-Neto, Grasiele~B. Santos, and Ward Struyve.
\newblock {Quantum-to-classical transition of primordial cosmological
  perturbations in de~Broglie-Bohm quantum theory: the bouncing scenario}.
\newblock \emph{Physical Review D}, 89:\penalty0 023517, 2014.
\newblock gr-qc: 1309.2670.

\bibitem[{D.N. Spergel et al.}(2007)]{WMAP2007}
{D.N. Spergel et al.}
\newblock {Three-Year Wilkinson Microwave Anisotropy Probe (WMAP) Observations:
  Implications for Cosmology}.
\newblock \emph{Astrophysical Journal Supplement Series}, 107:\penalty0 377,
  2007.
\newblock astro-ph: 0603449.

\bibitem[Roser(2015{\natexlab{b}})]{Roser2015CosmExtension}
Philipp Roser.
\newblock {An extension of cosmological dynamics with York time}.
\newblock 2015{\natexlab{b}}.
\newblock gr-qc: 1407.4005.

\bibitem[Thiemann(2008)]{Thiemann2008}
Thomas Thiemann.
\newblock \emph{{Modern Canonical Quantum General Relativity}}.
\newblock Cambridge University Press, 2008.

\bibitem[Peter et~al.(2007)Peter, Pinho, and
  Pinto-Neto]{PeterPinhoPintoneto2007}
Patrick Peter, E.~Pinho, and Nelson Pinto-Neto.
\newblock A non inflationary model with scale invariant cosmological
  perturbations.
\newblock \emph{Physical Review D}, 75, 2007.
\newblock hep-th: 0610205.

\bibitem[Peter and Pinto-Neto(2008)]{PeterPintoNeto2008}
Patrick Peter and Nelson Pinto-Neto.
\newblock {Cosmology without Inflation}.
\newblock \emph{Physical Review D}, 78:\penalty0 063506, 2008.
\newblock gr-qc: 0809.2022.

\bibitem[Valentini(2010)]{Valentini2010InflCosm}
Antony Valentini.
\newblock {Inflationary Cosmology as a Probe of Primordial Quantum Mechanics}.
\newblock \emph{Physical Review D}, 82:\penalty0 063513, 2010.

\bibitem[Valentini and Colin(2015)]{ValentiniColin2015a}
Antony Valentini and Samuel Colin.
\newblock {Primordial quantum nonequilibrium and large-scale cosmic anomalies}.
\newblock \emph{Physical Review D}, 92:\penalty0 043520, 2015.
\newblock astro-ph: 1407.8262.

\bibitem[Valentini(2015)]{Valentini2015a}
Antony Valentini.
\newblock {Statistical anisotropy and cosmological quantum relaxation}.
\newblock 2015.
\newblock astro-ph: 1510.02523.

\bibitem[Underwood and Valentini(2015)]{UnderwoodValentini2015}
Nicolas Underwood and Antony Valentini.
\newblock {Quantum field theory of relic nonequilibrium systems}.
\newblock \emph{Physical Review D}, 92:\penalty0 063531, 2015.
\newblock hep-th: 1409.6817.

\bibitem[John(2015)]{John2015}
Moncy~V. John.
\newblock {Exact Classical Correspondence in Quantum Cosmology}.
\newblock \emph{Gravitation and Cosmology}, 21:\penalty0 208, 2015.
\newblock gr-qc: 1405.7957.

\bibitem[Feynman and Hibbs(1965)]{FeynmanHibbs1965}
R.P. Feynman and A.R. Hibbs.
\newblock \emph{{Quantum Mechanics and Path Integrals}}.
\newblock McGraw-Hill, 1965.

\bibitem[Bell(1987)]{Bell1987}
J.S. Bell.
\newblock \emph{Speakable and Unspeakable in Quantum Mechanics}.
\newblock Cambridge University Press, 1987.

\bibitem[Bender(2005)]{BenderIntro2005}
Carl Bender.
\newblock {Introduction to PT-symmetric quantum theory}.
\newblock \emph{Contemporary Physics}, 46:\penalty0 277--292, 2005.
\newblock quant-ph: 0501052.

\end{thebibliography}

%\begin{thebibliography}{99}
%\small

%\end{thebibliography}

\end{document}